# X-WACoDa: An XML-based approach for Warehousing and Analyzing Complex Data

**Hadj Mahboubi**
**Jean-Christian Ralaivao**
**Sabine Loudcher**
**Omar Boussaïd**
**Fadila Bentayeb**
**Jérôme Darmont**

*Université de Lyon (ERIC Lyon 2)*
*5 avenue Pierre Mendès-France*
*69676 Bron Cedex*
*France*
E-mail: *first_name.last_name@univ-lyon2.fr*

**ABSTRACT**
Data warehousing and OLAP applications must nowadays handle complex data that are not only numerical or symbolic. The XML language is well-suited to logically and physically represent complex data. However, its usage induces new theoretical and practical challenges at the modeling, storage and analysis levels; and a new trend toward XML warehousing has been emerging for a couple of years. Unfortunately, no standard XML data warehouse architecture emerges. In this paper, we propose a unified XML warehouse reference model that synthesizes and enhances related work, and fits into a global XML warehousing and analysis approach we have developed. We also present a software platform that is based on this model, as well as a case study that illustrates its usage.

# INTRODUCTION

Data warehouses form the basis of decision-support systems (DSSs). They help integrate production data and support On-Line Analytical Processing (OLAP) or data mining. These technologies are nowadays mature. However, in most cases, the studied activity is materialized by numeric and symbolic data, whereas data exploited in decision processes are more and more diverse and heterogeneous. The development of the Web and the proliferation of multimedia documents have indeed greatly contributed to the emergence of data that can:
- be represented in various formats (databases, texts, images, sounds, videos...);
- be diversely structured (relational databases, XML documents...);
- originate from several different sources;
- be described through several channels or points of view (a video and a text that describe the same meteorological phenomenon, data expressed in different scales or languages...);
- change in terms of definition or value over time (temporal databases, periodical
- surveys...).

We term data that fall in several of the above categories *complex data* (Darmont *et al.*, 2005). For example, analyzing medical data regarding high-level athletes has lead us to jointly exploit information under various forms: patient records (classical database), medical history (text), radiographies and echographies (multimedia documents), physician diagnoses (texts or audio recordings), etc. (Darmont & Olivier, 2006; Darmont & Olivier, 2008)

Managing such data involves lots of different issues regarding their structure, storage and processing (Darmont & Boussaïd, 2006); and classical data warehouse architectures must be reconsidered to handle them. The XML language (Bray *et al.*, 2006) bears many interesting features for representing complex data (Boussaïd *et al.*, 2007; Boussaïd *et al.*, 2008; Darmont *et al.*, 2003; Darmont *et al.*, 2005). First, it allows embedding data and their schema, either implicitly, or explicitly through schema definition. This type of metadata representation suits data warehouses very well. Furthermore, we can benefit from the semi-structured data model's flexibility, extensibility and richness. XML document storage may be achieved either in relational, XML-compatible Database Management Systems (DBMSs) or in XML-native DBMSs. Finally, XML query languages such as XQuery (Boag *et al.*, 2007) help formulate analytical queries that would be difficult to express in a relational system (Beyer *et al.* 2004; Beyer *et al.*, 2005). In consequence, there has been a clear trend toward XML warehousing for a couple of years (Baril & Bellahsène, 2003; Hümmer *et al.*, 2003; Nassis *et al.*, 2005; Park *et al.*, 2005; Pkorný, 2002; Vrdoljak *et al.*, 2003; Zhang *et al.*,2005).

Our own motivation is to handle complex data into a complete decision-support process, which requires their integration and representation under a form processable by on-line analysis and/or data mining techniques (Darmont *et al.*, 2003). We have already proposed a full, generic data warehousing and on-line analysis process that includes two broad axes (Boussaïd *et al.*, 2008):
- data warehousing, including complex data integration and modeling;
- complex data analysis.

More precisely, the approach we propose consists in representing complex data as XML documents. Then, we recommend an additional layer to prepare them for analysis. Complex data under the form of XML documents are thus multidimensionally modeled to obtain an XML data warehouse. Finally, complex data analysis can take place from this warehouse, with on-line analysis, data mining or a combination of the two approaches.

Unfortunately, in this context, though XML and XQuery are normalized, XML DBMSs and data warehouse architectures are not. The XML warehouse models and approaches proposed in the literature share a lot of concepts (originating from classical data warehousing), but they are nonetheless all different. In this paper, we aim at addressing this issue. We first quite exhaustively present and discuss related work regarding XML warehousing and OLAP. Then, we motivate and recall our XML warehousing methodology, where XML is used for integrating and warehousing complex data for analysis. Our main contribution is a unified, reference XML data warehouse architecture that synthesizes and enhances existing models. We also present an XML warehousing software platform that is architectured around our reference model and illustrate its usage with a case study. Finally, we conclude this paper and provide future research directions.

# RELATED WORK

In this section, we first recall a couple of fundamental definitions before detailing literature work related to XML data warehousing and OLAP.

## Definitions

A *data warehouse* is a copy of transaction data specifically structured for query and analysis (Kimball, 2002). More formally, a data warehouse is a subject-oriented, integrated, time-variant and non-volatile collection of data in support of management's decision making process (Inmon, 2005). A single-subject data warehouse is typically referred to as a *datamart*, while data warehouses are generally enterprise in scope (Reed *et al.*, 2007).

A *star schema* is the simplest data warehouse schema. Shaped like a star, it consists of a single, central *fact table* linked to peripheral *dimensions*. A *fact* represents a business measurement (Kimball, 2002) and is constituted of references to its descriptive dimensions and a set of usually numeric and additive *measures* (such as sale amounts, for instance). Dimensions contain textual descriptors (*members*) of the studied business (Kimball, 2002). Star-like schema including hierarchies in dimensions (e.g., town, region and country granularity levels in a geographical dimension) are termed *snowflake schemas*. Star-modeled data warehouses with several fact tables are termed *constellation schemas*.

Finally, *On-Line Analytical Processing* or OLAP (Codd *et al.*, 1994) is an approach for efficiently processing decision-support, analytical queries that are multidimensional by nature. Data are stored in multidimensional arrays called *data cubes* that are typically extracted from data warehouses. Data cubes are then manipulated with the help of OLAP operators.

## XML data warehousing

Several studies address the issue of designing and building XML data warehouses. They propose to use XML documents to manage or represent facts and dimensions. The main objective of these approaches is to enable a native storage of the warehouse and its easy interrogation with XML query languages.

Research in this area may be subdivided into three families. The first family particularly focuses on Web data integration for decision-support purposes. However, actual XML warehouse models are not very elaborate. The second family of XML warehousing approaches is explicitly based on classical warehouse logical models (star-like schemas). They are used when dimensions are dynamic and they allow the support of end-user analytical tools. Finally, the third family we identify relates to document warehousing. It is based on user-driven approaches that are applied when an organization has fixed warehousing requirements. Such requirements correspond to typical or predictable results expected from an XML document warehouse or frequent user-query patterns.

### XML Web warehouses

The objective of these approaches is to gather XMLWeb sources and integrate them into a Web warehouse. Vrdoljak *et al.* (2003) introduce the design of aWeb warehouse that originates from XML Schemas describing operational sources. This method consists in preprocessing XML Schemas, in creating and transforming the schema graph, in selecting facts and in creating a logical schema that validates a data warehouse.

Golfareli *et al.* (2001) also propose a semi-automatic approach for building a datamart conceptual schema from XML sources. The authors show how data warehouse multidimensional design may be carried out starting directly from XML sources. They also propose an algorithm that solves the problem of correctly inferring the information needed for data warehousing.

Finally, the designers of the Xyleme system propose a dynamic warehouse for XML data from the Web that supports query evaluation, change control and data integration (Xyleme, 2001). No particular warehouse model is proposed, though.

### XML data warehouses

In his XML-star schema, Pokorný (2002) models a star schema in XML by defining dimension hierarchies as sets of logically connected collections of XML data, and facts as XML data elements.

Hümmer *et al.* (2003) propose a family of templates enabling the description of a multidimensional structure (dimension and fact data) for integrating several data warehouses into a virtual or federated warehouse. These templates, collectively named XCube, consist of three kinds of XML documents with respect to specific schemas: *XCubeSchema* stores metadata; *XCubeDimension* describes dimensions and their hierarchical levels; and *XCubeFact* stores facts, i.e., measures and the corresponding dimension references. These federated templates are not actually directly related to XML warehousing, but they can definitely be used for representing XML star schemas.

Rusu *et al.* (2005) propose a methodology, based on the XQuery technology, for building XML data warehouses. This methodology covers processes such as data cleaning, summarization, intermediating XML documents, updating/linking existing documents and creating fact tables. Facts and dimensions are represented by XML documents built with XQueries.

Park *et al.* (2005) introduce a framework for the multidimensional analysis of XML documents, named XML-OLAP. XML-OLAP is based on an XML warehouse where every fact and dimension is stored as an XML document. The proposed model features a single repository of XML documents for facts and multiple repositories of XML documents for dimensions (one repository per dimension).

Eventually, Boussaïd *et al.* (2006) propose an XML-based methodology, named XWarehousing, for warehousing complex data. They use XML Schema as a modelling language to represent user analysis needs. These needs are then compared to complex data stored in heterogeneous XML sources. Information needed for building an XML cube is extracted from these sources and transformed into OLAP facts.

Note that all these studies, though all different, more or less converge toward a unified XML warehouse model. They mostly differ in the way dimensions are handled and the number of XML documents that are used to store facts and dimensions.

### XML document warehouses

Nassis *et al.* (2005) propose a conceptual approach for designing and building an XML repository, named xFACT. They exploit object-oriented concepts and propose to select dimensions based on user requirements. To enhance the XML data warehouse's expressiveness, these dimensions are represented by XML virtual views. In this approach, the authors assume that all dimensions are part of fact data and that each fact is described in a single XML document.

Rajugan *et al.* (2005) also propose a view-driven approach for modeling and designing an XML fact repository, named GxFact. GxFact gathers xFACTs (distributed XML warehouses and datamarts) in a global company setting. The authors also provide three design strategies for building and managing GxFact to model further hierarchical dimensions and/or global document warehouses.

Baril and Bellahsène (2003) envisage XML data warehouses as collections of views represented by XML documents. Views, defined in the warehouse, allow to filter and to restructure XML sources. A warehouse is defined as a set of materialized views and provides a mediated schema that constitutes a uniform interface for querying the XML data warehouse. Following this approach, Baril and Bellahsène have developed a system named DAWAX.

Finally, Zhang *et al.* (2005) propose an approach to materialize XML data warehouses based on frequent query patterns discovered from historical queries. The authors apply a hierarchical clustering technique to merge these queries and therefore build the warehouse.

## Multidimensional analysis over XML data

Though several studies from the literature address the issue of XML data warehousing, fewer actually push through the whole decision-support process and address the multidimensional analysis of XML data. To query XML cubes, Park *et al.* (2005) propose a multidimensional expression language, XML-MDX. The authors supplement the Microsoft multidimensional expression language, MDX, with two additional statements: `CREATE XQ-CUBE` to create XML cubes, and `SELECT` to query them. In addition, the authors define seven aggregation operators: `ADD`, `LIST`, `COUNT`, `SUMMARY`, `TOPIC`, `TOP KEYWORD` and `CLUSTER`. Some operators are inherited from the relational context, while others are designed for non-additive data and exploit text mining techniques.

Beyer *et al.* (2005) argue that analytical queries written in XQuery are difficult to read, write, and process efficiently. To address these issues, the authors propose to extend XQuery FLWOR expressions with an explicit syntax for grouping and numbering query results. They also present solutions dealing with the homogeneous and hierarchical aspect of XML data for explicit grouping problems.

In the same context, Wang *et al.* (2005) present concepts for XOLAP (OLAP on XML data). The authors define a general XML aggregation operator, GXaggregation. This operator permits property extraction from dimensions and measures through their XPath expression. Hence, computing statistics over XML data becomes more flexible. This process is performed with functions that aggregate heterogeneous data over hierarchies. The authors also envisage to embed GXaggregation in an XML query language such as XQuery.

Finally, Ben Messaoud *et al.* (2006a) propose an OLAP aggregation operator that is based on an automatic clustering method: OpAC. The authors' proposal enables precise analyses and provides semantic aggregates for complex data represented by XML documents. OpAC has been applied onto XML cubes output by the XWarehousing approach (Boussaïd *et al.*, 2006).

## XML WAREHOUSING AND ANALYSIS METHODOLOGY

In a data warehousing process, the data integration phase is crucial. Data integration is a hard task that involves reconciliation at various levels (data models, data schemas, data instances, semantics). Nowadays, in most organizations, XML documents are becoming a casual way to represent and store data. Therefore, new efforts are needed to integrate XML in classical business applications. Integrating heterogeneous and complex information in DSSs requires special consideration. Existing ETL (Extract, Transform, Load) tools that organize data into a common syntax are indeed ill-adapted to complex data. Furthermore, if XML documents must be prepared for future OLAP analyses, storing them in a data repository is not enough. Through these documents, a more interesting abstraction level, completely oriented toward analysis objectives, must be expressed. It is thus necessary to structure XML data with respect to a data warehouse multidimensional reference model.

Though feeding data warehouses with XML documents is getting increasingly common, methodological issues arise. The multidimensional organization of data warehouses is indeed quite different from the semi-structured organization of XML documents. A data warehouse architecture is subject-oriented, integrated, consistent, and data are regularly refreshed to represent temporal evolutions. Then, how can multidimensional design be carried out with a semi-structured formalism such as XML?

XML may be characterized by two aspects. On one hand, it helps store and exchange data through XML documents. On the other hand, XML Schemas are relevant for describing data. Multidimensional modeling helps structure data for query and analysis. An XML formalism can thus be used to describe the various elements of a multidimensional model (Boussaïd *et al.*, 2006). But XML can only be considered as a logical and physical description tool for future analysis tasks on complex data. The reference conceptual model remains the star schema and its derivatives.

One challenge we address in our approach is to propose a multidimensional model (thus oriented for analysis) that is described in XML, to derive a physical organization of XML documents that contributes to performance enhancement. To support this choice, we propose a modeling process (Figure 1) that achieves complex data integration (Boussaïd *et al.*, 2003; Boussaïd *et al.*, 2007; Boussaïd *et al.*, 2008).

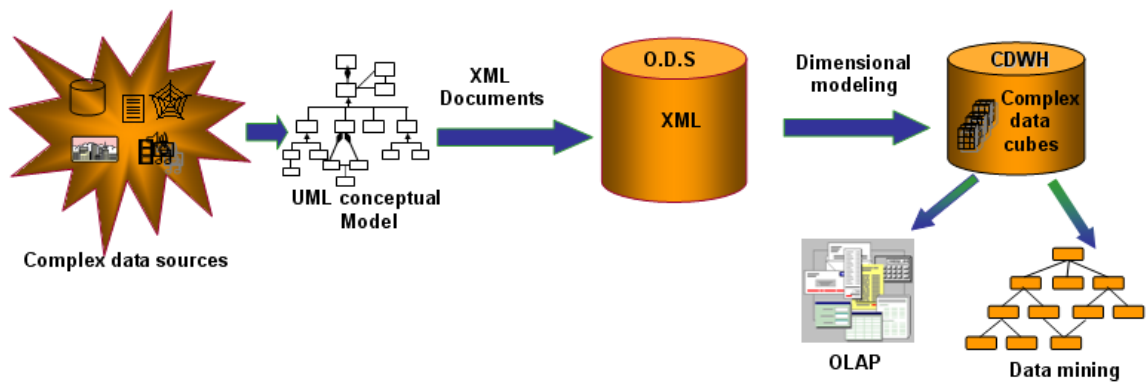

*Figure 1 XML data warehousing and analysis process*

We first design a conceptual UML model for a complex object. This UML model is then directly translated into an XML Schema, which we view as a logical model. At the physical level, XML documents that are valid against this logical model may be mapped into a relational, object-relational or XML-native database. In this paper, we focus on the latter family of DBMSs. After representing complex data as XML documents, we physically integrate them into an Operational Data Store (ODS), which is a buffer ahead of the actual warehouse.

At this stage, it is already possible to mine the stored XML documents directly, e.g., with XML structure mining techniques. In addition, to further analyze these documents' contents efficiently, it is interesting to warehouse them, i.e., devise a multidimensional model that allows OLAP analyses. However, classical OLAP tools are ill-adapted to deal with complex data. OLAP facts representing complex data indeed require appropriate tools and aggregation methods to be analyzed. A new idea consists in combining OLAP and data mining algorithms to provide new OLAP operators able to compute more significant aggregates, mainly on complex data cubes. In this context, we have proposed three approaches.

1. *Sparse data visualization:* with multiple correspondence analysis, we have reduced the negative effect of sparsity by reorganizing cube cells (BenMessaoud *et al.*, 2006).
2. *Complex fact aggregation:* with agglomerative hierarchical clustering, we obtain aggregates that are semantically richer than those provided by traditional multidimensional structures (BenMessaoud *et al.*, 2006a).
3. *Explanation of possible relationships in multidimensional data:* we have designed a new algorithm for guided association rule mining in data cubes (BenMessaoud *et al.*, 2007).

## MODELING XML DATA WAREHOUSES

In this section, we mainly present our analysis of related work and our proposal to unify and enhance existing XML warehouse models. Our reference model notably exploits the concept of virtual key reference we define previously.

### Preliminary definitions

An XML document is defined as a labeled graph (an *XML graph*) whose nodes represent document elements or attributes, and whose edges represent the element-subelement (or parent-child) relationship. Edges are labeled with element or attribute names. Let $E$ be the set of distinct element names, $A$ the set of distinct attribute names and $V$ the set of element and attribute values.

An XML graph can be denoted by the expression: $\Gamma := \langle t, l, \psi \rangle$, where $t$ is an finite ordered tree, $l$ a function that labels a node in $t$ with symbols from $E \cup A$, and $\psi$ □ a function that associates a node in $t$ to its corresponding value in $V$. The root node of $t$ is denoted $root_t$.

In an XML graph, a node $e$ can be referenced by another node $e'$. Let $a$ and $a'$ be two attribute nodes that are children of $e$ et $e'$, respectively. Then, $e$ references $e'$ if and only if $\psi(a) = \psi(a')$ and $l(a)$ 6= $l(a')$. Such a link is referred to as a *virtual key reference*.

### XML data warehouse reference model

Previous XML data warehouse architectures converge toward a unified model. They mostly differ in the way dimensions are handled and in the number of XML documents that are used to store facts and dimensions. We may distinguish four different families of physical architectures:

1. one XML document for storing facts and another for storing all dimension related information (XCube);
2. a collection of XML documents that each embed one fact and its related dimensions (X-Warehousing);
3. a collection of XML documents where facts and dimensions are each stored in one separate document (XML-OLAP);
4. one XML document for storing facts and one XML document for storing each dimension (analogous to relational star-like schemas).

A performance evaluation study of these different representations has been performed by Boukraa et *al.* (2006). The authors have built four XML mammographic warehouses with respect to the four architectures we have enumerated. These XML warehouses have been stored on the eXist XML-native DBMS (Meier, 2002). Then, the authors measured the response time of a fixed XQuery workload over these warehouses and showed that representing facts in one single XML document and each dimension in one XML document allowed the best performance. They especially conclude that storing facts in one single XML document helps decrease their scanning cost.

Moreover, this representation also allows to model constellation schemas without duplicating dimension information. Several fact documents can indeed share the same dimensions. Also, since each dimension and its hierarchical levels are stored in one XML document, dimension updates are more easily and efficiently performed than if dimensions were either embedded with the facts or all stored in one single document.

Hence, we adopt the architecture model from Figure 2 to represent our XML data warehouse. It is actually the translation of a classical snowflake schema from the relational model to the XML model, i.e., tables become XML documents. Note, however, that XML warehouses can bear irregular structures that are not possible in relational warehouses. On the physical level, our reference data warehouse is composed of the following XML documents:
- *dw-model.xml* represents the warehousemetadata, basically the warehouse schema: fact document(s) structure (related dimensions, numeric measures), dimension documents structure (including any hierarchical level information, each level being characterized by descriptive member attributes);
- a set of *facts$_f$.xml* documents that each help store information related to set of facts $f$, i.e., dimension references and measure values;
- a set of *dimension$_d$.xml* documents that each help store a given dimension $d$'s member values.

Figure 3 represents the *dw-model.xml* document's graph structure. The equivalent XML Schema definition we actually use in practice is available on-line to accommodate space constraints[1]. The *dw-model.xml* document defines the multidimensional structure of the warehouse. Its root node, *DW-model*, is composed of two types of nodes: *dimension* and *FactDoc* nodes. A *dimension* node defines one dimension, its possible hierarchical levels (*Level* elements) and attributes (including their types), as well as the path to the corresponding *dimension$_d$.xml* document. A *FactDoc* element defines a fact, i.e., its measures, virtual key references to the corresponding dimensions, and the path to the corresponding *facts$_f$.xml* document.

---

[1] http://eric.univ-lyon2.fr/~hmahboubi/X-WACoDa/Schemas/dw-model.xsd

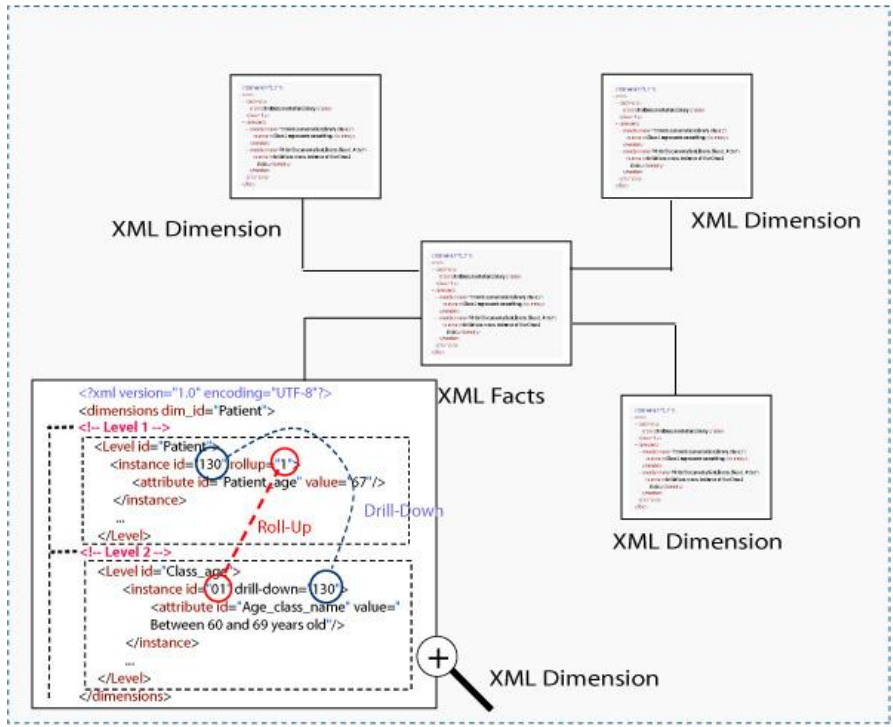

*Figure 2 XML data warehouse model*

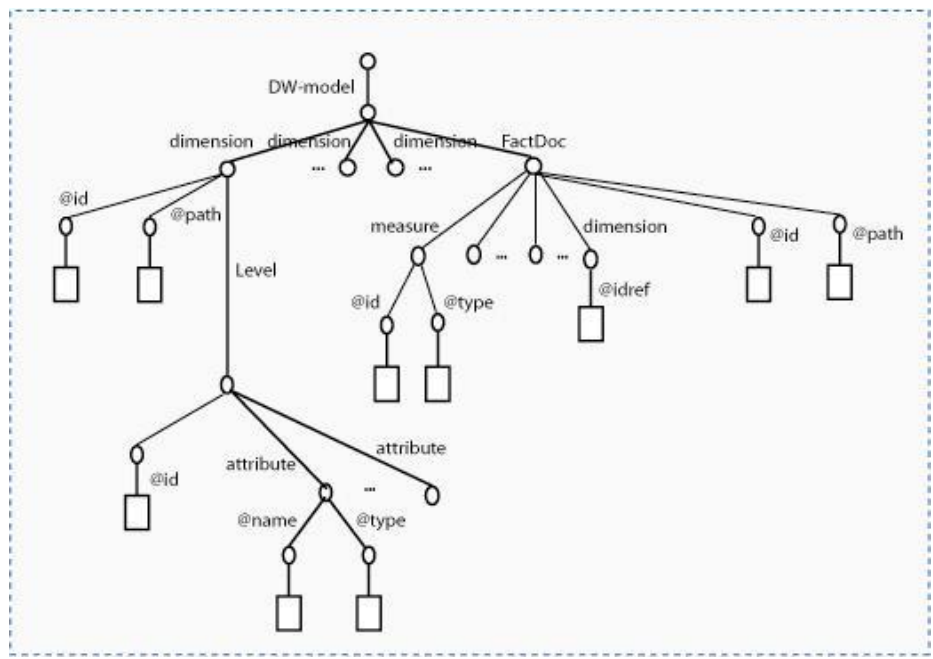

*Figure 3 dw-model.xml graph structure*

Figure 4(a) represents the *facts$_f$.xml* documents' graph structure. Its equivalent XML Schema is available on-line[2]. A *facts$_f$.xml* document stores facts. The document root node, *FactDoc*, is composed of *fact* subelements that each instantiate a fact, i.e., measure values and dimension virtual key references. These identifier-based references support the fact-to-dimension relationship.

Finally, Figure 4(b) represents the *dimension$_d$.xml* documents' graph structure. Its equivalent XML Schema is available on-line[3]. A *dimension$_d$.xml* document helps instantiate one dimension, including any hierarchical level. Its root node, *dimension*, is composed of *Level* nodes. Each one defines a hierarchy level composed of *instance* nodes that each define the level's member attribute values. In addition, an *instance* element contains *Roll-up* and *Drill-Down* attributes that define the hierarchical relationship within dimension *d*. More precisely, these multivalued virtual key references help link sets of instances of a given dimension hierarchy level to their aggregate in the next level (*Roll-up*), and vice-versa (*Drill-Down*).

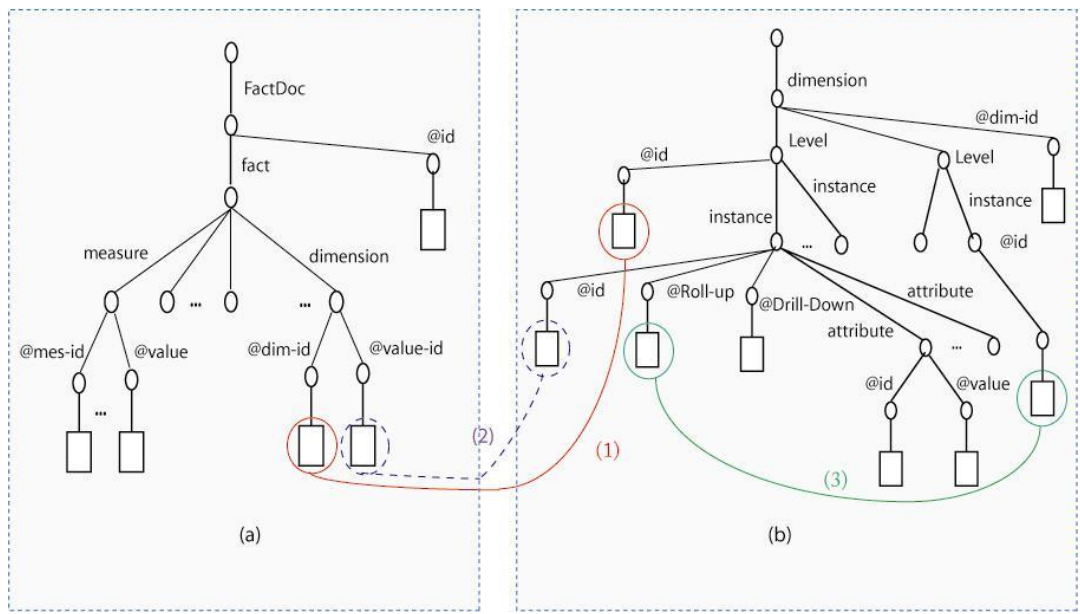

*Figure 4 facts$_f$.xml (a) and dimension$_d$.xml (b) graph structure*

## SOFTWARE PLATFORM

In order to support experimentations and projects that validate our XML complex data warehousing approach (X-WACoDa), we have developed a software platform we also named X-WACoDa by extension. We detail in this section its architecture, as well as a case study that illustrates our whole approach and exploits the software platform.

X-WACoDa's architecture is represented in Figure 5. It consists of three components that are further detailed in the following sections:
1. an ETL component that allows to extract complex data representations in an homogeneous XML format and to integrate them into the XML data warehouse. This component is based on user requirements;

---

[2] http://eric.univ-lyon2.fr/~hmahboubi/X-WACoDa/Schemas/facts.xsd

[3] http://eric.univ-lyon2.fr/~hmahboubi/X-WACoDa/Schemas/dimension.xsd

2. an actual warehouse component that manages XML data with respect to our XML data warehouse reference model;
3. an analysis component that permits various analyses (OLAP, data mining...) over the data warehouse.

**ETL component**

The ETL component in X-WACoDa is actually a set of tools that form a chain of data extraction, transformation and loading into the XML data warehouse.

First, SMAIDoC is a multi-agent system that allows complex data extraction and transformation into a predefined XML format (Boussaïd *et al.*, 2003). SMAIDoC (Figure 6) consists of a set of agents that collect or extract data (*DataAgent*), structure and model them (*WrapperAgent*), translate them and generate the corresponding XML documents (*XML Creator*), and finally store the obtained XML documents into a database (*XML2RDBAgent*).

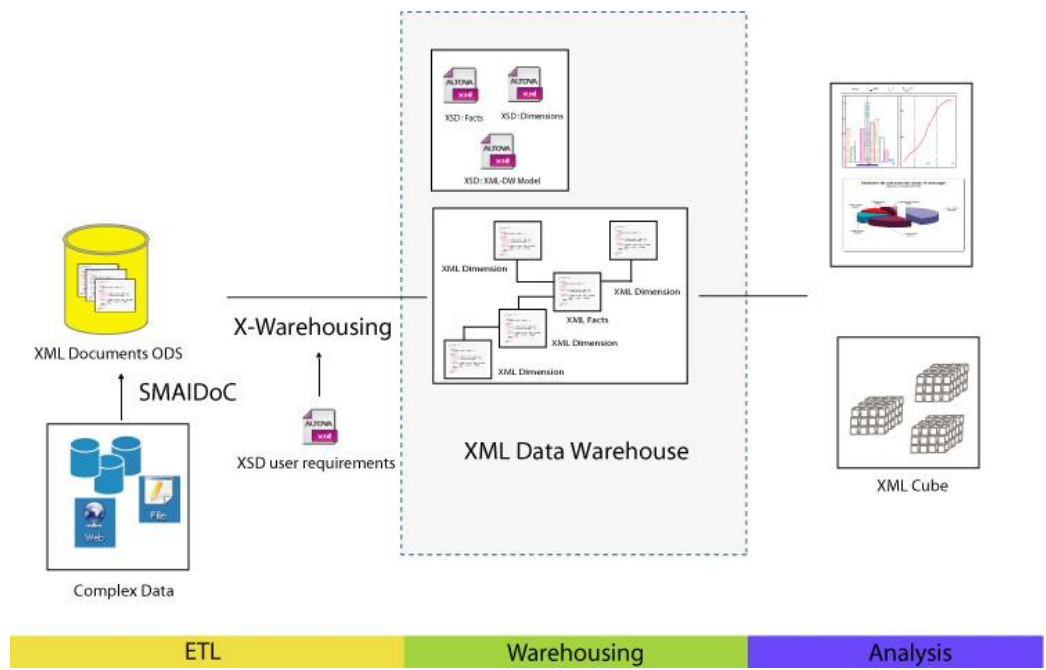

*Figure 5 X-WACoDa software platform architecture*

Then, UML2XML is a graphical interface that helps users express their analysis requirements in UML and outputs an XML document representing the warehouse schema: *dw-model.xml*.

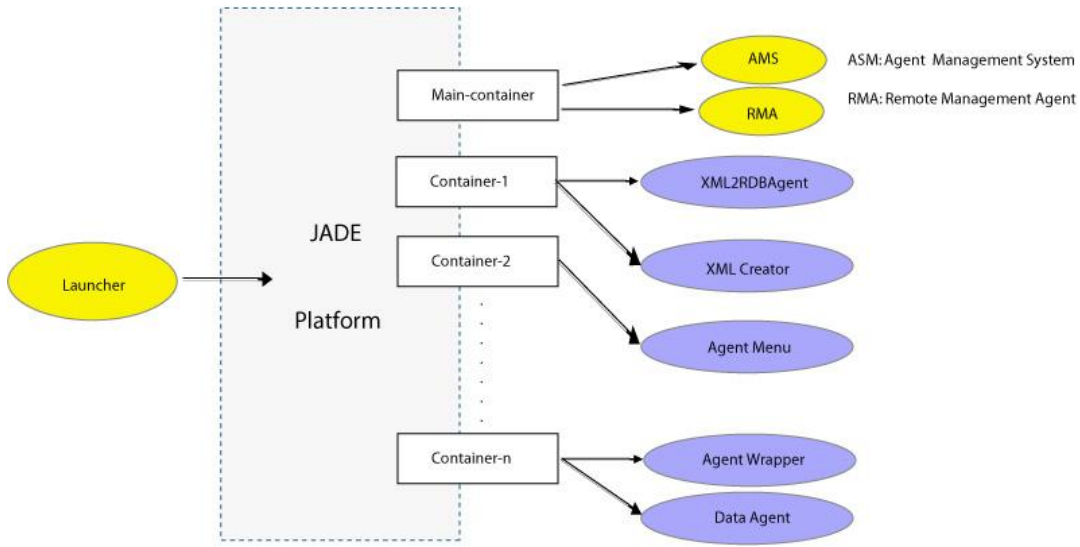

*Figure 6 SMAIDoC architecture*

Finally, X-Warehousing generates the warehouse's XML documents (Boussaïd *et al.*, 2006). XWarehousing inputs the complex data that have been formated in XML by SMAIDoC, as well as the user analysis requirements produced by UML2XML. Its architecture is made of two modules (Figure 7):
1. a *loader module* that loads user requirements and XML data source schemas, and then transforms them into attributes trees (Golfarelli *et al.*, 2001);
2. a *merger module* that merges these attribute trees into one final attribute tree representing the warehouse schema. The merging process operates through the application of fusion functions (pruning and grafting).

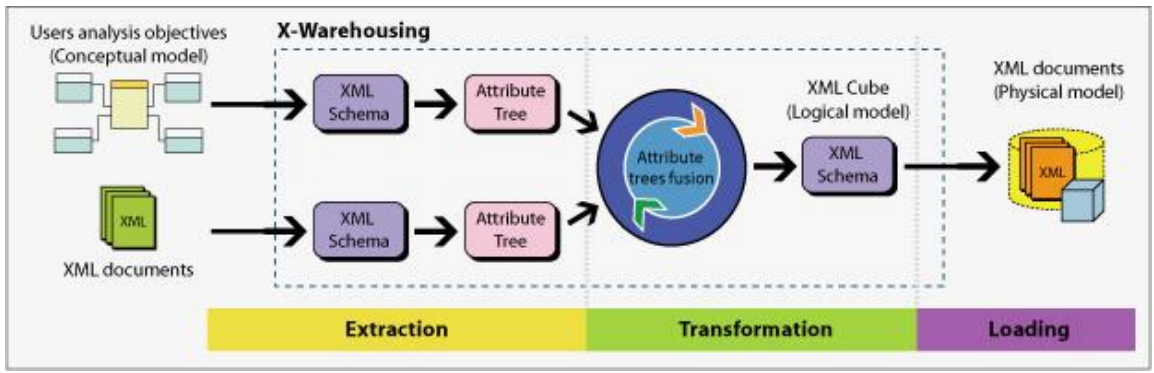

*Figure 7 X-Warehousing architecture*

**Data warehouse**

X-Warehousing directly outputs XML documents that form an XML data warehouse with respect to our reference model. Fact data are stored in one or several *facts$_f$.xml* documents (several in the case of a constellation schema). Dimension data are stored, for each dimension *d*, in the *dimension$_d$.xml* document. Data exploitation is achieved through a DBMS. Two types of DBMSs currently support XML data storage, management, and query processing and optimization: relational, XML-enabled systems that map XML data into relational tables, and XML-native DBMSs.

In our approach, we focus on XML-native systems. They indeed consider XML documents as the fundamental unit of storage. They define a specific XML storage schema for XML documents, and store and retrieve them according to this model, which includes elements, attributes, their contents and order. Moreover, many XML databases provide a logical model for grouping XML documents, called collections or libraries. XML-native DBMSs also implement XML query engines supporting XPath and XQuery. In addition, some XML-native DBMSs support the XML:DB Application Programming Interface (API) that features a form of implementation independent access to XML data.

Query engine performance is the primary criterion when selecting an XML native DBMS. A good candidate system must be capable to perform complex queries in reasonable response times and to store a large volumes of data. This may currently be seen as a weak point in our choice, since XML-native DBMSs are not mature yet, but we are also working in parallel on optimizing their performances (Mahboubi *et al.*, 2006; Mahboubi *et al.*, 2008; Mahboubi *et al.*, 2008a).

**Analysis component**

X-WACoDa's analysis component is constituted of two subcomponents. First, an *ad hoc* reporting application helps users create specific and customized decision support queries. This application is based on XML:DB and allows database connection, sending and saving XML query results in XML format. Analytical queries are expressed in XQuery. We selected XQuery because it allows performing complex queries over multiple XML documents. In addition, we extended XQuery's FLWOR clauses with an explicit grouping construct comparable to the SQL group by clause, in order to allow common business analysis (i.e., OLAP-like) queries (Mahboubi *et al.*, 2006).

The second analysis subcomponent, MiningCubes, is a Web-based application that includes a set of on-line analysis and mining components (BenMessaoud *et al.*, 2006a). Analysis components aim at loading data and performing multidimensional explorations (through two or three-dimension views). Data mining components implement methods such as agglomerative hierarchical clustering or frequent itemset mining. From a user's point of view, MiningCubes integrates these components in a transparent way. For instance, a factorial approach can be used to represent and reorganize relevant OLAP facts, association rules may be mined from an OLAP cube, or clustering may be exploited to aggregate non-additive dimension members in roll-up/drill-down operations.

**Case study**

Let us now apply the X-WACoDa approach onto a real-world application domain and consider complex data from the Digital Database for ScreeningMammography[4] (DDSM). DDSM gathers 2604 medical history cases of anonymous patients. Each case contains an ASCII text file representing general information about a patient and four LJPEG radiography image files. These data are issued from multiple sources and encoded through different file types, and may thus be considered complex.

The first step in our approach is to transform DDSM data in XML format to guarantee an homogeneous representation and to allow data integration into an XML data warehouse. Such XML documents individually describe one whole medical case by gathering study information (date, examination...), patient information, and radiography image descriptors (file name, url, scanner resolution...). All documents bear the same structure and are valid against an XML Schema. In this case study, we selected and processed 1406 XML documents.

---

[4] http://marathon.csee.usf.edu/Mammography/Database.html

The second step in our approach is to design a *dw-model.xml* document representing user analysis requirements, with the help of the UML2XML software. In the present case study, this document represents a star schema composed of *Suspicious region* facts (suspected cancerous regions) characterized by the *Region length* and *Number of regions* measures. *dw-model.xml* also describes dimensions and their hierarchies. We obtain ten dimensions: *Patient, Lesion type, Assessment, Subtlety, Pathology, Date of study, Date of digitization, Digitizer, Scanner image* and *Boundary*.

In a third step, both the XML documents representing complex medical data and *dw-model.xml* are submitted to the X-warehousing software to actually build an XML data warehouse. X-warehousing outputs ten XML documents representing dimensions: *Patient.xml, Lesion_type.xml, Assessment.xml, Subtlety.xml, Pathology.xml, Date_of_study.xml, Date_of_digitization.xml, Digitizer.xml, Scanner image.xml* and *Boundary.xml*; and one XML documents containing facts: *facts.xml*. These documents constitute the XML data warehouse. We chose to store this warehouse within the X-Hive XML-native DBMS[5]. X-Hive allows the native storage of large documents and supports XQuery. It also provides APIs for storing, querying, retrieving, transforming and publishing XML data.

Finally, our analysis application exploits a set of decision-support queries expressed in XQuery. Figure 10 provides an example of analytical query that returns the total number of suspicious regions for fifty-eight-year-old patients. It performs one join operation between the *Patient* dimension and facts, a selection and an aggregation operation. Variable *q* stores the *Number of regions* measure values used by the aggregation function.

```
for $x in //FactDoc/fact
$y in //dimenion/Level[@id='Patient']/instance
let $q := $x/measure[@mes-id='Number_of_regions']/@value
where $y/attribute/@id='Patient_age'
and $a/attribute/@value='58'
and $x/dimension/@value-id=$y/@id
and $x/dimension/@dim-id='Patient'
return fn:sum($q)
```

*Figure 8 Sample XQuery over the DDSM XML warehouse*

## CONCLUSION AND PERSPECTIVES

Nowadays, data processed by DSSs tend to be more and more complex and pose new challenges to the data warehousing community. To efficiently manage and analyze complex data, we propose a full, generic, XML-based data warehousing and on-line analysis approach: X-WACoDa. This approach includes complex data integration, multidimensional modeling and analysis. In this paper, we identified some substantial heterogeneity in XML warehouse models from the literature, and thus focused on proposing a unified reference XML data warehouse architecture. We also presented a software platform, also named X-WACoDa, which implements our ideas.

Research perspectives in the young field of XML warehousing are numerous. Regarding complex data integration, we aim at extracting useful knowledge for warehousing from data themselves, by applying data mining techniques. We plan to study a metadata representation of data mining results in mixed structures combining XML Schema and the Resource Description Framework (RDF). These description languages are indeed well-suited for expressing semantic properties and relationships between metadata. On a technical level, SMAIDoC could also be extended to converse with on-line search engines (including some semantic Web tools) and exploit their answers to identify and qualify Web data sources.

---

[5] http://www.x-hive.com/products/db/

Our choice of an "all-XML" architecture also leads to address performance problems. Our research in this area shows that using indexes and/or materialized views significantly improves response time for typical analytical queries expressed in XQuery (Mahboubi *et al.*, 2006; Mahboubi *et al.*, 2008; Mahboubi *et al.*, 2008a). Further gains in performance can be achieved, though. For instance, it is widely acknowledged that indexes and materialized views are mutually beneficial to each other. We have designed a method for simultaneously selecting indexes and materialized views in the relational context, which we aim at adapting to the XML context. Finally, our performance optimization strategies could be better integrated in a host XML-native DBMS. In particular, the mechanism for rewriting queries exploiting materialized views would be more efficient if it was part of the system.

We finally have a couple of perspectives regarding complex data analysis. First, since the structure of XML documents carries some relevant information, we plan to exploit XML structure mining to, e.g., discover tag relevance. Relevant tags may then be selected as measures or dimensions in our multidimensional modeling process. Second, we have underlined with the MiningCubes software the benefit of associating OLAP and data mining to enhance on-line analysis power. We are currently working on extending on-line analysis with new capabilities such as explanation and prediction, with the aim of better handling data complexity.

## REFERENCES

Baril, X. & Bellahsène, Z., (2003). Designing and Managing an XML Warehouse. XML Data Management: Native XML and XML-enabled Database Systems. *Addison Wesley*, 455–473.

BenMessaoud, R., Boussaïd, O. & Loudcher-Rabaséda, S., (2006). Efficient Multidimensional Data Representation Based on Multiple Correspondence Analysis. *In: ACM SIGKDD International Conference on Knowledge Discovery and Data Mining (KDD'06), Philadelphia, USA*. ACM Press. 662-667.

BenMessaoud, R., Boussaïd, O. & Loudcher-Rabaséda, S., (2007). A multiple correspondence analysis to organize data cubes. *Vol. 155(1) of Databases and Information Systems IV – Frontiers in Artificial Intelligence and Applications*. IOS Press, 133–146.

BenMessaoud, R., Loudcher-Rabaséda, S. & Boussaïd, O. (2006a). A Data Mining-Based OLAP Aggregation of Complex Data: Application on XML Document. *International Journal of Data Warehousing & Mining 2 (4)*, 1–26.

Beyer, K. S., Chamberlin, D. D., Colby, L. S., Ozcan, F., Pirahesh, H. & Xu, Y. (2005). Extending XQuery for Analytics. *In: ACM SIGMOD International Conference on Management of Data (SIGMOD'05), Baltimore, USA*. ACM, 503–514.

Beyer, K. S., Cochrane, R., Colby, L. S., Ozcan, F. & Pirahesh, H., (2004). XQuery for Analytics: Challenges and Requirements. *In: First InternationalWorkshop on XQuery Implementation, Experience and Perspectives <XIME-P/>, Paris, France*, 3–8.

Boag, S., Chamberlin, D., Fernandez, M., Florescu, D., Robie, J. & Simeon, J., (2007). XQuery 1.0: An XML Query Language. W3C, http://www.w3.org/TR/xquery/.

Boukraa, D., BenMessaoud, R., Boussaïd & O., (2006). Proposition d'un Modèle Physique pour les Entrepôts XML. *In: Atelier Systèmes Décisionnels (ASD'06), 9th Maghrebian Conference on Information Technologies (MCSEAI'06), Agadir, Morocco.*

Boussaïd, O., BenMessaoud, R., Choquet, R. & Anthoard, S., (2006). X-Warehousing: An XML-Based Approach for Warehousing Complex Data. *In: 10th East-European Conference on Advances in Databases and Information Systems (ADBIS'06), Thessaloniki, Greece*. Vol. 4152 of Lecture Notes in Computer Science. Springer, 39–54.

Boussaïd, O., Bentayeb, F. & Darmont, J., (2003). A Multi-Agent System-Based ETL Approach for Complex Data. *In: 10th ISPE International Conference on Concurrent Engineering: Research and Applications (CE'03), Madeira, Portugal*. 49–52.

Boussaïd, O., Darmont, J., Bentayeb, F. & Loudcher-Rabaseda, S., (2008). Warehousing complex data from the Web. *International Journal of Web Engineering and Technology 4(4).* 408-433 (Invited paper).

Boussaïd, O., Tanasescu, A., Bentayeb, F. & Darmont, J., (2007). Integration and dimensional modelling approaches for complex data warehousing. *Journal of Global Optimization 37 (4)*, 571–591.

Bray, T., Paoli, J., Sperberg-McQueen, C., Maler, E., Yergeau, F. & Cowan, J., (2006). Extensible Markup Language (XML) 1.1. *W3C, 2nd Edition*, http://www.w3.org/TR/2006/REC-xml11-20060816/.

Codd, E., Codd, S. & Salley, C., (1994). Providing OLAP (On-line Analytical Processing) to User-Analysts: An IT Mandate. White paper, E.F. Codd Associates.

Darmont, J. & Boussaïd, O. (Eds.), 2006. Managing and Processing Complex Data for Decision Support. Idea Group Publishing.

Darmont, J., Boussaïd, O., Bentayeb, F. & Sabine Loudcher-Rabaseda, Y. Z., (2003). Web multiform data structuring for warehousing. *Vol. 22 of Multimedia Systems and Applications.* Kluwer Academic Publishers, 179–194.

Darmont, J., Boussaïd, O., Ralaivao, J.-C. & Aouiche, K., (2005). An architecture framework for complex data warehouses. *In: 7th International Conference on Enterprise Information Systems (ICEIS'05), Miami, USA*. INSTICC, 370–373.

Darmont, J. & Olivier, E., (2006). A complex data warehouse for personalized, anticipative medicine. *In: 17th Information Resources Management Association International Conference (IRMA'06),Washington, USA.* Idea Group Publishing, 685–687.

Darmont, J. & Olivier, E., (2008). Biomedical Data Warehouses. *Encyclopaedia of Healthcare Information Systems*. IGI-Global, to appear.

Golfarelli, M., Rizzi, S. & Vrdoljak, B., (2001). Data Warehouse Design from XML Sources. *In: 4th International Workshop on Data Warehousing and OLAP (DOLAP'01), Atlanta, USA.* ACM Press, 40–47.

Hümmer, W., Bauer, A., Harde & G., (2003). XCube: XML for data warehouses. *In: 6th International Workshop on Data Warehousing and OLAP (DOLAP'03), New Orleans, USA.* ACM, 33–40.

Inmon, W., (2005). Building the Data Warehouse, 4th Edition. John Wiley & Sons.

Kimball, R., Ross, M., (2002). The Data Warehouse Toolkit, 2nd Edition. John Wiley & Sons.

Mahboubi, H., Aouiche, K. & Darmont, J., (2006). Materialized View Selection by Query Clustering in XML Data Warehouses. *In: 4$^{th}$ International Multiconference on Computer Science and Information Technology (CSIT'06), Amman, Jordan*. 68–77.

Mahboubi, H., Aouiche, K. & Darmont, J., (2008). A Join Index for XML Data Warehouses. *In: 2008 International Conference on Information Resources Management (Conf-IRM'08), Niagara Falls, Canada.*

Mahboubi, H. & Darmont, J., (2007). Indices in XML databases. *Encyclopedia of Database Technologies and Applications, second edition.* IGI-Global, to appear.

Meier, W., (2002). eXist: An Open Source Native XML Database. In: Web, Web-Services, and Database Systems, NODe 2002 Web and Database-Related Workshops, Erfurt, Germany. Vol. 2593 of Lecture Notes in Computer Science. Springer, 169–183.

Nassis, V., Rajugan, R., Dillon, T. S. & Rahayu, J.W., (2004). Conceptual Design of XML DocumentWarehouses. *In: 6th International Conference on Data Warehousing and Knowledge Discovery (DaWaK'04), Zaragoza, Spain.* Vol. 3181 of Lecture Notes in Computer Science. Springer, pp. 1–14.

Nassis, V., Rajugan, R., Dillon, T. S. & Rahayu, J. W., (2005). A Requirement Engineering Approach for Designing XML-View Driven, XML Document Warehouses. *In: 29th International Conference on Computer Software and Applications (COMPSAC'05), 25-28 July 2005, Edinburgh, UK.* IEEE Computer Society, pp. 388–395.

Nassis, V., Rajugan, R., Dillon, T. S. & Rahayu, J. W., (2005). Conceptual and Systematic Design Approach for XML Document Warehouses. *International Journal of Data Warehousing & Mining 1 (3)*, 63–86.

Park, B.-K., Han, H. & Song, I.-Y., (2005). XML-OLAP: A Multidimensional Analysis Framework for XML Warehouses. *In: 7th International Conference on Data Warehousing and Knowledge Discovery (DaWaK'05), Copenhagen, Denmark.* Vol. 3589 of Lecture Notes in Computer Science. Springer, pp. 32–42.

Pokorný, J., (2002). XML Data Warehouse: Modelling and Querying. *In: 5th International Baltic Conference (BalticDB&IS'06), Tallin, Estonia.* Institute of Cybernetics at Tallin Technical University, pp. 267–280.

Rajugan, R., Chang, E. & Dillon, T. S., (2005). Conceptual Design of an XML FACT Repository for Dispersed XML Document Warehouses and XML Marts. *In: 20th International Conference on Computer and Information Technology (CIT'05), Shanghai, China.* IEEE Computer Society, pp. 141–149.

Reed, M., (2007). A definition of data warehousing. Intranet Journal: http://www.intranetjournal.com/features/datawarehousing.html.

Rusu, L. I., Rahayu, J. W., Taniar & D., (2005). A Methodology for Building XML Data Warehouse. *International Journal of Data Warehousing & Mining 1 (2),* 67–92.

Vrdoljak, B., Banek, M. & Rizzi, S., (2003). DesigningWebWarehouses from XML Schemas. *In: 5th International Conference on DataWarehousing and Knowledge Discovery (DaWaK'03), Prague, Czech Republic.* Vol. 2737 of Lecture Notes in Computer Science. Springer, pp. 89–98.

Wang, H., Li, J., He, Z. & Gao, H., (2005). OLAP for XML Data. *In: 1st International Conference on Computer and Information Technology (CIT'05), Shanghai, China.* IEEE Computer Society, pp. 233–237.

Xyleme, L., (2001). Xyleme: A Dynamic Warehouse for XML Data of the Web. *In: International Database Engineering & Applications Symposium (IDEAS'01), Grenoble, France.* IEEE Computer Society, pp. 3–7.

Zhang, J., Wang, W., Liu, H. & Zhang, S., (2005). X-warehouse: building query pattern-driven data. *In: 14th international conference on World Wide Web (WWW'05), Chiba, Japan.* ACM, pp. 896–897.

**KEY TERMS & DEFINITIONS**

Database management system (DBMS): Software set that handles structuring, storage, maintenance, update and querying of data stored in a database.

XML-enabled DBMS: Database system in which XML data may be stored and queried from relational tables. Such a DBMS must either map XML data into relations and translate queries into SQL, or implement a middleware layer allowing native XML storing and querying.

XML-native DBMS (NXD): Database system in which XML data are natively stored and queried as XML documents. An NXD provides XML schema storage and implements an XML query engine (typically supporting XPath and XQuery). eXist (Meier, 2002) and X-Hive (X-Hive Corporation, 2007) are examples of NXDs.

XML data warehouse: XML database that is specifically modeled (i.e., multidimensionally, with a star-like schema) to support XML decision-support and analytic queries.

Web warehouse: "a shared information repository. A web warehouse acts as an information server that supports information gathering and provides value added services, such as transcoding, personalization." (Cheng et al., 2000)

XML document warehouse: an XML document repository dedicated to e-business and Web data analysis.

Complex data: data that present several axes of complexity for analysis, e.g., data represented in various formats, diversely structured, from several sources, described through several points of view, and/or versioned.